\documentclass{aa}  

\usepackage{graphicx}
\usepackage{txfonts}
\usepackage{hyperref}
\usepackage{xcolor}
\usepackage{ulem}

\begin{document}

   \title{Evolution of eccentric high-mass X-ray binaries. \\
          The case of GX~301--2}


   \author{Adolfo Simaz Bunzel\thanks{Fellow of CONICET}
          \inst{1,2}
          \and
          Federico Garc\'ia
          \inst{1,2}
          \and
          Jorge A. Combi
          \inst{1,2,3}
          \and
          Sylvain Chaty
          \inst{4}
  }

   \institute{
   Instituto Argentino de Radioastronom\'ia (CCT La Plata, CONICET; CICPBA; UNLP), C.C.5, (1894) Villa Elisa, Buenos Aires, Argentina
   \and
   Facultad de Ciencias Astron\'omicas y Geof\'{\i}sicas, Universidad Nacional de La Plata, Paseo del Bosque, B1900FWA La Plata, Argentina
   \and
   Departamento de Ingeniería Mecánica y Minera (EPSJ), Universidad de Jaén, Campus Las Lagunillas s/n Ed. A3, E-23071 Jaén, Spain
   \and
   Universit\'e Paris Cit\'e, CNRS, Astroparticule et Cosmologie, F-75013 Paris, France
   }

   \date{received ... 2022; accepted ... 2022}

 
  \abstract
  {
    The formation of neutron stars is associated with powerful astrophysical transients such as supernovae. In many cases, asymmetries in
    the supernova explosions are thought to be responsible for the large observed velocities of neutron stars.
  }
  {
    We aim to study the complete evolutionary history of one particular eccentric high-mass X-ray binary containing a neutron star,
    GX~301--2, and characterize the natal kick at the time of neutron star formation.
  }
  {
    We used the publicly available stellar-evolution code {\tt MESA} to evolve binaries from their initial stages until the core-collapse
    scenario. We incorporated a natal kick distribution based on observations to continue the evolution during the X-ray binary phase and
    search for candidates matching current observations of GX~301--2.
  }
  {
    We find that the range of initial masses is constrained to be less than around $30$~M$_\odot$ depending on the initial mass ratio, as
    higher initial masses will most likely end up producing a black hole. In the completely conservative mass-transfer scenario under
    study, only is an interaction between the stars when the donor is still burning Hydrogen in its core, the so-called Case A of
    mass transfer, able to produce progenitors for GX~301--2. The natal kick study favours kicks of variable strength, which in turn
    increases the tilt angle between the orbital angular momentum and the spin of the neutron star.
  }
  {
    We conclude that only a narrow initial progenitor parameter space is able to produce a binary such as GX~301--2 when assuming a
    completely conservative mass transfer. Additionally, the strength of the natal kick can span a wide range of values, but it can be
    constrained when considering new data concerning the systemic velocity of the binary. Finally, we derive the fraction of the expected
    number of binaries such as GX~301--2 in the Galaxy to be $\sim 6 \times 10^{-5}$, implying a really low chance of finding a binary
    similar to GX~301--2.
  }


   \maketitle

\section{Introduction}

The growing number of observations of Galactic X-ray binaries (XRBs) provides tools that help us to constrain the formation of compact
objects and related processes during which the kinematic properties of these binaries are expected to change. One interesting, yet
unresolved issue, is the process under which natal kicks are imparted to the compact object, either a black hole (BH) or a neutron star
(NS), during its formation, associated with supernova explosions. These natal kicks are believed to be responsible for the large transverse
motions on the plane of the sky of NSs in binaries and isolated pulsars. Although there is no direct observation of such kicks, their
existence can be inferred from the peculiar velocities measured in a variety of isolated NSs, and also from the high eccentricity observed
in some XRBs containing NSs. Gathering observational information about the current state of XRBs, such as masses, orbital periods, and
positions in the Hertzsprung$-$Russell (HR) diagram, can then be complemented with kinematic constraints to discern the complete
evolutionary path and to provide a consistent picture of compact object formation scenarios and the crucial role played by natal kicks.

Of particular interest is GX~301--2, an XRB discovered by \citet{1973ApJ...184..237R} that contains a compact object and an optical star,
named Wray~977 \citep[BR~Cru,][]{1973ApJ...186L..81V,1977Natur.269...21B}. Based on spectroscopic observations, \citet{1995A&A...300..446K}
classified Wray~977 as a B1~Ia+ hypergiant, which makes it the only hypergiant star known to be within a binary system. In addition, the
mass function derived from pulse-timing analyses indicates that the mass of Wray~977 is larger than $31$~M$_\odot$
\citep{1986ApJ...304..241S}, making it a member of the high-mass X-ray binary (HMXB) class. \citet{2013ApJ...764..185C} derived a
distance to the binary of $3.10 \pm 0.64$~kpc using near-infrared magnitudes from the Two Micron All Sky Survey (2MASS) point-source
catalogue. The source presents X-ray pulsations with a period of 11.6 minutes \citep{1976ApJ...209L.119W} and a history of rapid spin-up
episodes \citep{1997ApJ...479..933K}. Studying the occurrence of X-ray flares, \citet{1982MNRAS.199..915W} determined an orbital period
of~$41.5$~days. Furthermore, GX~301--2 is one of the most eccentric systems among all the HMXBs known \citep{2006A&A...455.1165L}. In the
period between 2018 and 2020, the binary experienced a new spin-up process \citep{2021MNRAS.506.2712D}. Observations during this period
suggest a disc-accretion scenario favouring the efficient transfer of angular momentum \citep{2021MNRAS.504.2493L}, rather than a
wind-accretion one. The spin orientation of the NS has been under discussion recently with some authors favouring a retrogradely spinning
NS \citep{2020MNRAS.494.2178M}, while others tend to favour a prograde spin solution \citep{2020MNRAS.496.3991L}. Given the highly unequal
masses measured in the binary components, as well as their particular nature, GX~301--2 stands as a very rare and unique object, whose
individual properties are particularly different from every other known XRB, which naturally makes its rate of occurrence quite low, as we
subsequently address and discuss in this paper.

The combination of the individual masses (and the mass ratio) and orbital parameters in this binary make it an interesting source to try
and understand how binary evolution leads to core collapse and, in addition, what the role being played by the natal kicks being imparted
onto newly born compact objects is. In one of the first detailed calculations of its evolution, \citet{1996ApJ...463..297B} favoured a
very massive NS progenitor, of $\sim 45$~M$_\odot$, originating from the initially most massive star in the binary system. Along the same
lines, \citet{1998A&A...331L..29E} proposed a massive progenitor of $\gtrsim 50$~M$_\odot$, by assuming a fully inefficient mass-transfer
(MT) scenario. Perhaps the most detailed binary evolution study on the progenitor of GX~301--2 is presented in \citet{1999A&A...350..148W}.
In their work, the authors explore a channel that would arise from the smallest possible initial mass for the progenitor binary, with the
aim being to find the lower limit in masses leading to a BH. For that reason, they restricted their search to almost equal initial masses
in a close binary such that stars went through an efficient MT phase during the main sequence (MS) evolution, known as Case~A of MT
\citep{1967ZA.....65..251K}, while avoiding the inclusion of a kick onto the NS. They found that an initial primary mass of
$\sim 26$~M$_\odot$ with a companion of similar characteristics in a circular orbit with a period of $\sim 3.5$~days could match the
observed properties of GX~301--2. More recently, \citet{2012arXiv1208.2422B} studied a similar case to that in \citet{1999A&A...350..148W},
but focussing on the future evolution of the binary. They predict that a MT phase should occur after the hypergiant companion swells enough
and, given the unequal mass ratio, this MT phase will most likely be dynamically unstable, leading to a common-envelope (CE) phase
\citep{2013A&ARv..21...59I}, which even with the most optimistic assumptions will be unable to eject its envelope and survive this phase,
thus finishing its evolution as a peculiar object in a CE merger \citep{1977ApJ...212..832T,1998ApJ...502L...9F,2001ApJ...550..357Z}.

In this paper we follow the study presented in \citet{1999A&A...350..148W} by using updated physical prescriptions for important stellar
evolutionary phases. In addition, we address the role of supernova-induced kicks on the NS in light of the new measurements done by the
{\it Gaia} satellite on the proper motion of the binary \citep{2021A&A...649A...1G} and its radial velocity \citep{2006A&A...457..595K}.
The paper is organized as follows. Section~\ref{section:model} contains a brief description of the setup of our detailed binary evolution
models. In Section~\ref{section:results} we start by presenting a representative evolution of a binary leading to conditions similar to
current GX~301--2 properties. We then show the results from a grid of detailed binary simulations of stars until core collapse. We then
present results on the impact of natal kicks in Section~\ref{section:kicks}. Our concluding remarks are finally given in
Section~\ref{section:conclusions}.

\section{Methods: Stellar evolution}
\label{section:model}

Interacting  binaries are calculated using the detailed stellar-evolution code {\tt MESA}
\citep[version 10398,][]{Paxton2011,Paxton2013,Paxton2015,Paxton2018,Paxton2019}. All computed stars are assumed to be non-rotating and to
have an initial solar metallicity content, $Z = Z_\odot = 0.017$ \citep{1998SSRv...85..161G}. We used default nuclear reaction networks
present in {\tt MESA}: {\tt basic.net}, {\tt co\_burn.net}, and {\tt approx21.net} which were switched dynamically once later stages in the
evolution were reached. Convective regions, which were obtained from the Ledoux criterion \citep{1947ApJ...105..305L}, were modelled using
the standard mixing-length theory \citep{1958ZA.....46..108B} with a mixing-length parameter $\alpha_{\rm MLT} = 2.0$. Semi-convection was
modelled following \citet{1983A&A...126..207L} with an efficient parameter $\alpha_{\rm SC} = 1$. We included an exponential overshooting
beyond convective boundaries \citep{2000A&A...360..952H} with overshoot mixing parameters of $f = 0.01$ and $f_0 = 0.005$. In order to
avoid some numerical problems during the evolution of massive stars until the Wolf-Rayet (WR) stage, we used the MLT++ formalism as
presented in \citet{Paxton2013} which reduces the super-adiabaticity in regions where the convective velocities approach the speed of
sound. Additionally, the effect of thermohaline mixing follows \citet{1980A&A....91..175K} with an efficiency parameter of
$\alpha_{\rm th} = 1$. The modelling of stellar winds follows that of \citet{2011A&A...530A.115B} as described in
\citet{2017A&A...604A..55M}.

Binary systems are initially assumed to have circular orbits with both components being at their zero-age main sequence (ZAMS). Binary
interaction was computed with the {\tt MESAbinary} module of {\tt MESA}. When one of the stars in the binary overflowed its Roche lobe, the
MT rate was implicitly computed following the prescription of \citet{1990A&A...236..385K}. Furthermore, we assume there is
no mass lost during this phase, that is, a fully conservative case of MT throughout the entire evolution. As late stages in the evolution
do not produce significant changes to the binary properties since the remaining time-to-core collapse is notably short, we assumed the
occurrence of a supernova (SN) when one of the stars reached core Carbon depletion, leaving behind a compact remnant. The mass of the
remnant was obtained from the {\tt delayed} prescription from \citet{2012ApJ...749...91F}, which depends on the masses of the Carbon-Oxygen
(CO) core and the envelope at core Carbon depletion\footnote{Input files with the chosen parameters that are necessary to reproduce our
results are available on Zenodo: \url{https://zenodo.org/record/7261481}.}.

\section{Results}
\label{section:results}

\subsection{An example evolution}
\label{subsection:example-evolution}

\begin{figure}
    \centering
    \includegraphics[width=\hsize]{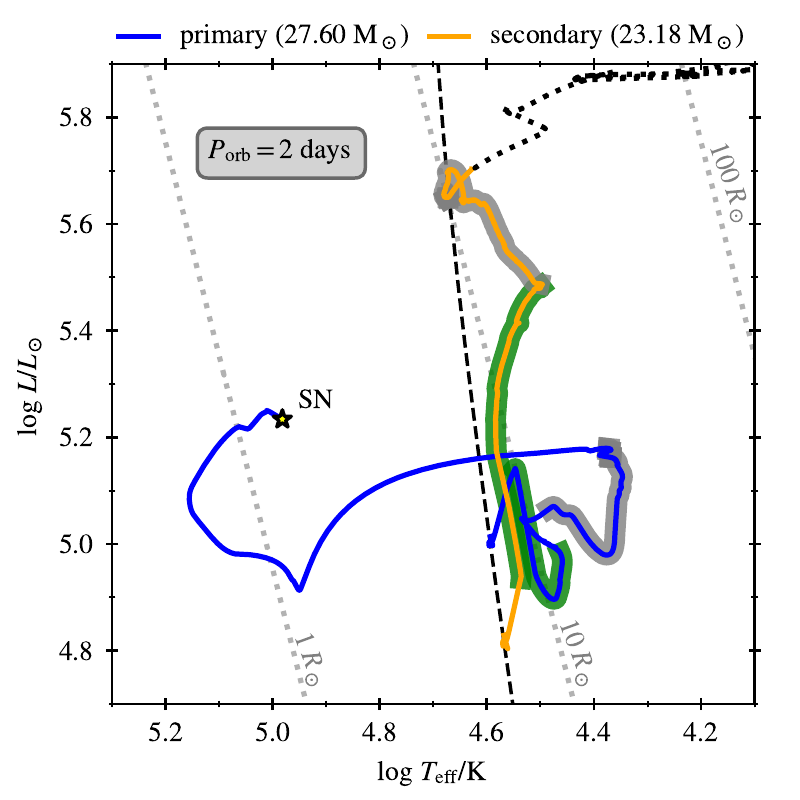}
    \caption{Evolutionary tracks in the HR diagram of a close binary system with an initial primary mass of $27.60$~M$_\odot$ and a
      secondary mass of $23.18$~M$_\odot$. This binary is assumed to start its evolution in a circular orbit with $P_{\rm orb} = 2$~days,
      in a fully conservative MT scenario throughout its entire lifetime. The evolution of the primary (secondary) star is represented in
      solid blue (orange) line. Different MT phases are plotted on top of each star: in green (grey) we show Case~A (early Case~B) MT. The
      dotted grey lines represent constant radii for 1, 10, and 100~R$_\odot$, while the dotted black line designates the evolution of the
      secondary star after the formation of the first compact object (GX~301--2) if the companion NS star was absent. The yellow star marks
      the location where the primary exploded as a SN. For comparison, the ZAMS corresponding to stars of the same composition is shown by
      the black dashed line.}
    \label{fig:hr_exemplary_model}
\end{figure}

Before examining the wider parameter space explored, we describe here a representative case of binary evolution, which is followed by the
majority of the binary systems modelled in this work. In Figure~\ref{fig:hr_exemplary_model} we show the evolutionary tracks in the HR
diagram of a primary star of~$27.60$~M$_\odot$ with a companion (secondary) star of $23.18$~M$_\odot$ (which corresponds to a mass ratio
$q = 0.84$) in an initially circular orbit with an orbital period of $P_{\rm orb} = 2$~days. We consider that both stars are born in the
ZAMS. During core Hydrogen (H) burning, the primary overflows its Roche lobe and a Case~A of MT starts \citep{1967ZA.....65..251K}. This
phase, which happens in a nuclear timescale, strips the primary of a fraction of its envelope mass and stops shortly before the primary
depletes the H in its core. During this MT phase, the mass ratio is reversed and the secondary becomes the more massive of the pair. The
secondary star, which is accreting all the mass that is transferred, becomes overluminous while keeping an almost constant effective
temperature, T$_{\rm eff}$.

While the primary is crossing the Hertzsprung gap (HG), it overflows its Roche lobe for a second time and an early Case~B of MT phase
starts \citep{1967ZA.....65..251K}. In this fast MT episode, most of the H-rich envelope of the primary is removed, until detachment is
reached. Although a small fraction of envelope mass remains in the primary at the end of this second MT phase, strong winds from the WR
phase help to remove it. By the end of its evolution, the primary star, which is almost completely depleted of H, explodes as a hot He-rich
star, leaving behind a $2.2$~M$_\odot$~NS. The mass of the compact remnant was calculated following the prescription of
\citet{2012ApJ...749...91F}, as we mention later in Section \ref{subsec:grid}. On the other hand, the secondary, which was rejuvenated by
the accreted H-rich material of the previous MT episodes, is still burning H in its core.

\begin{table}[]
\caption{Binary parameters of GX~301--2 and stellar parameters of Wray~977 corresponding to the values listed in
    \citet{2006A&A...457..595K}: the mass of Wray~977 $M_{\rm opt}$, the mass of the compact object companion $M_{\rm X}$, the orbital
    period $P_{\rm orb}$, the eccentricity $e$, as well as the radius, effective temperature, and luminosity of Wray~977 $R_{\rm opt}$,
    $T_{\rm eff}$, and $L$, respectively.}
\centering
\begin{tabular}{lr}
\hline
\hline
Parameter & Value \\
\hline
$M_{\rm opt}$~[M$_\odot$] &     $33-53$       \\
$M_{\rm X}$~[M$_\odot$]   &     $1.85-2.5$          \\
$P_{\rm orb}$~[days]      &     $41.498 \pm 0.002$  \\
$e$                       &     $0.462 \pm 0.014$   \\
\hline
$R_{\rm opt}$~[R$_\odot$] &     $70$                \\
$T_{\rm eff}$~[K]         &     $18\,100 \pm 500$   \\
$\log\,(L/L_\odot)$       &     $5.67$              \\
\hline
\hline
\end{tabular}
\label{table1:gx-parameters}
\end{table}

\subsection{Grid of models}
\label{subsec:grid}

Given the large difference in masses between the components of GX~301--2, we can set some contraints for the exploration of the
initial parameters: masses ($M_1$, $M_2$) and P$_{\rm orb}$. First, if we assume that the stars evolved as if they were isolated, that
is, with no MT interaction between them, then the progenitor of the NS should have been more massive than current measurements for
Wray~977, with an initial mass above $40$~M$_\odot$. However, masses as high as those are likely to contain a massive CO core, which would
end up collapsing into a BH instead of a NS \citep{2012ApJ...749...91F,2016ApJ...821...38S,2020MNRAS.499.2803P}. Second, the total initial
mass $M_{\rm tot,i}$ in the system must be larger than current estimated values, which imposes a lower limit in masses such that
$M_{\rm tot,i} \gtrsim 40$~M$_\odot$. From this, we suggest that the system likely experienced an interaction phase in which the initially
less massive star accreted a significant amount of mass from its companion. In order to maximize this stage of interaction so as to
increase the mass of the accreting star, we initialized the stars close to each other ensuring that the MT phase started when H was being
burned in their cores, as in \citet{1999A&A...350..148W}. Furthermore, to avoid an unstable MT phase leading to a merger of the stars, the
initial mass ratio $q$ should be close to unity  \citep{1987ApJ...318..794H}. Thus, for our binary models, we considered primary stars with
masses between $M_1 = 22-30$~M$_\odot$, mass ratios in the $q = 0.8-0.9$ range, and $P_{\rm orb} = 2-10$~days. For the simulation grid, we
used linear steps of~$0.2$~M$_\odot$, $0.02$ and $0.2$~days, respectively. This grid was chosen after performing an initially wider search
in all three mentioned parameters, aimed at obtaining masses consistent with GX~301--2, as listed in Table~\ref{table1:gx-parameters}.

\begin{figure}
    \centering
    \includegraphics[width=\hsize]{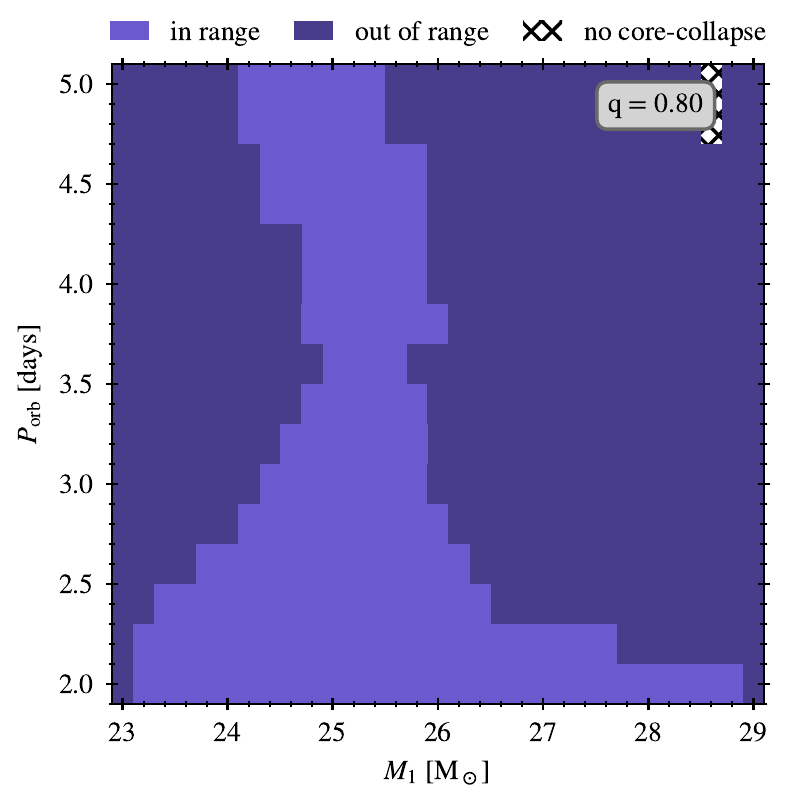}
    \caption{Example of a grid of binary systems showing initial periods (in days) and initial primary masses for a fixed initial
      mass ratio of $q = 0.8$. Each rectangle represents a single detailed binary evolution model. Light blue is used for binaries with a
      primary collapsing into a NS while having a companion mass within the values derived by \citet{2006A&A...457..595K},
      (named {'in range'}). In a darker blue colour, we show binaries with primaries forming BHs or being outside the constraint imposed
      by the observed masses (named {'out of range'}). The hatched region represents a binary that ends its evolution without
      reaching a core-collapse stage. For the entire grid of models computed, we find that these non-collapsing models either merge during
      a CE phase between two non-degenerate stars, or stop due to numerical issues. Other initial mass ratios produce similar results.}
    \label{fig:example_grid}
\end{figure}

In Figure~\ref{fig:example_grid} we show a grid of binary models explored for a fixed mass ratio of~$q=0.80$, where each rectangle is a
detailed binary calculation. As shown in Figure~\ref{fig:example_grid}, there is a narrow range of primary masses in which we find binaries
that end up with a star collapsing into a NS and a companion star with a mass within the values derived by \citet{2006A&A...457..595K}, as
presented in Table~\ref{table1:gx-parameters}, while having an initial orbital period between 2 and 5~days. In the lower end of primary
masses, we find that the allowed initial orbital periods become wider when increasing the initial masses. Between $24 - 26$~M$_\odot$, the
range of allowed initial orbital periods is the longest, with compatible masses obtained for $P_{\rm orb}= 2 - 5$~days. Additionally, we
find that for the higher initial masses, the range of initial orbital periods decreases to our lower limit of $P_{\rm orb} = 2$~days. Thus,
in order to obtain the proper mass range, we needed a lower limit for the primary star mass of~$23$~M$_\odot$ and an upper limit
of~$29$~M$_\odot$.

We find progenitors with consistent masses only if the initial orbital period is below $5$~days. This limiting value also depends on the
initial primary mass: an increment in the mass requires a shorter initial period. Moreover, binaries with orbital periods of less than
$2$~days fail to reach the core-collapse stage. In all of these cases, and as a consequence of the complete conservative MT regime
explored, both of the stars overflow their Roche lobe producing a contact system \citep{2006MNRAS.368.1742D,2015ApJ...812..102A}. These
binaries will likely end up in a CE phase \citep{2013A&ARv..21...59I} which, given the range of obtained orbital periods, will unavoidably
merge during the CE phase. Finally, this scenario is also found for binaries with orbital periods of more than $5$~days, in which the MT
leads to a contact system.

Binaries in which the primary star reaches the core Carbon depletion are assumed to collapse and produce either a NS or a BH according to
the conditions established by \citet{2012ApJ...749...91F} in the {\tt delayed} scenario. At this final stage, we find that those binaries
with masses {in range} of GX~301--2 are constrained to have pre-collapse primary masses between 5 and 7~M$_\odot$, companion stars of
34 to 39~M$_\odot$, and orbital periods in the range of 16 to 22~days.

\section{Natal kicks}
\label{section:kicks}

During an SN explosion, both mass loss and a kick imparted onto the newly formed NS change the orbital binary parameters. In order to
understand the impact that natal kicks have on the progenitor of GX~301--2, we selected the binary presented in
Section~\ref{subsection:example-evolution}, which presents components with masses at core collapse within the range derived for GX~301--2,
and we randomly drew~$2\,000$ kicks from a Maxwellian distribution with a root mean square of $\sigma = 265$~km~s$^{-1}$ in the magnitude
as constrained from pulsar observations \citep{2005MNRAS.360..974H}. We further assumed an isotropic orientation for the drawn kicks and we
reduced the kick strength by a factor of $(1 - f_{\rm fb})$, with $f_{\rm fb}$ being the fraction of mass that falls back to the proto-NS
at core collapse \citep{2012ApJ...749...91F}\footnote{We repeated this study by applying the same natal kick distribution to another
randomly drawn binary in the appropriate mass range in order to confirm the validity of these results.}. For all of these kicks, we
calculated the post-SN orbital period $P_{\rm post-kick}$ and eccentricity $e_{\rm post-kick}$ following \citet{1996ApJ...471..352K},
assuming no interaction between the ejecta of the SN and the companion of the exploding star. We then evolved the binaries that remained
bound after the imparted kick using {\tt MESA}, including the additions described in \citet{2021A&A...649A.114G}. Our goal is to find which
combinations of kick strengths and orientations are able to reproduce, after following the evolution from formation of the NS onwards,
current estimates for the binary parameters of GX~301--2. It is worth mentioning that among some of these additions, we included a
treatment for a putative CE phase. Given the extreme masses in GX~301--2, if at any time the companion star overflows its Roche lobe, the
development of a CE phase becomes unavoidable. Furthermore, given the high envelope binding energy, this CE phase finally leads to a merger
of the NS with the core of the companion star \citep{2012arXiv1208.2422B}.

\begin{figure}
    \centering
    \includegraphics[width=\hsize]{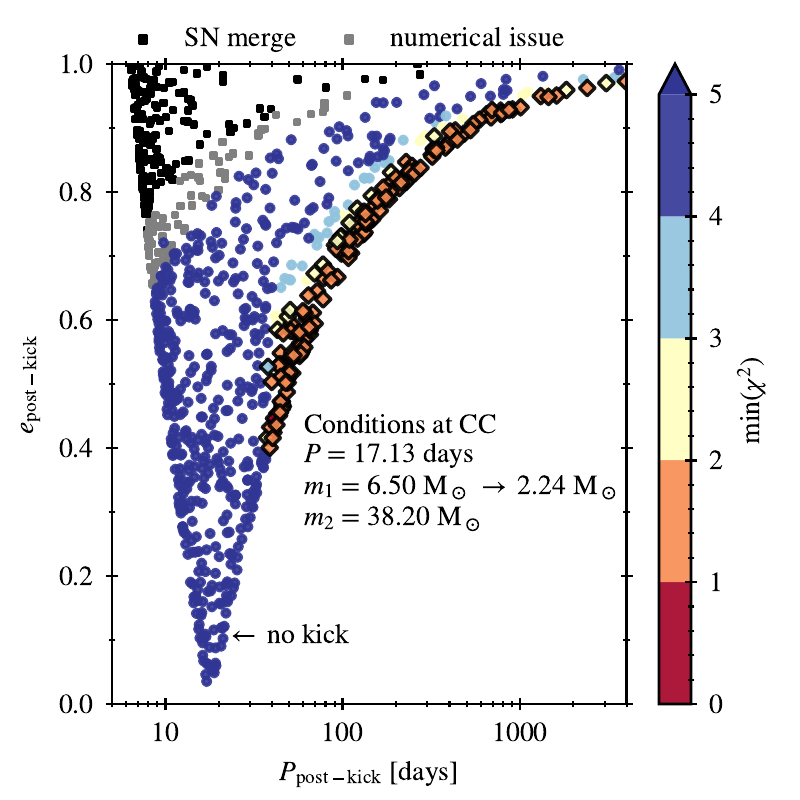}
    \caption{Post-kick binary systems in the period-eccentricity plane. Each point represents a detailed binary simulation between a
      non-degenerate donor star an a NS. The colours indicate the $\chi^2$ minimum found during the entire evolution of each binary,
      considering observational constraints on binary parameters (i.e. masses of the NS and its companion, orbital period, and
      eccentricity; see Table~\ref{table1:gx-parameters}). Diamonds represent binaries with matching binary properties to the observed
      ones. With black squares we show binaries merging just after core collapse due to the kick orientation, while grey squares represent
      binaries for which numerical problems were  encountered and the simulations were not completed. The location of the case of a
      symmetric kick in this plane is shown with an arrow.}
    \label{fig:progenitors}
\end{figure}

Figure~\ref{fig:progenitors} shows the outcome of all of the detailed simulations in the period-eccentricity plane. Orbital parameters can
only be found in a limited region derived from the different conservation laws \citep{1996ApJ...471..352K}. We see a clear relation between
binaries with numerical issues and those merging due to the orientation of the kick. This is experienced by binaries whose kicks lead to
systems in which the donor star overflows its Roche lobe and starts transferring mass just after the kick.

In order to constrain the orbital parameters (and the kicks associated with them), we analysed the evolution of each individual binary to
find the ones that best match the observed properties of GX~301--2. We quantified this by applying a minimum $\chi^2$ method to their
evolving individual masses, periods, and eccentricities, so that for each timestep taken in {\tt MESA} we could evaluate the corresponding
$\chi^2$ value. In order to do this, we imposed an uncertainty (systematic) in the orbital period of~0.5~days and of 0.1 in the
eccentricity for the values listed in Table~\ref{table1:gx-parameters}. These values replace the actual observational (statistical) errors,
since they are so small that they dominate the $\chi^2$ value. These best-matching binaries are shown as diamonds in
Figure~\ref{fig:progenitors}. We find compatible post-kick periods ranging from 40~days (with a respective $e_{\rm post-kick} \sim 0.4$) up
to 4000~days ($e_{\rm post-kick} \gtrsim 0.9$). These large ranges of $P_{\rm post-kick}$ and $e_{\rm post-kick}$ have been obtained as a
direct consequence of tides acting on the evolution of these binary parameters, which tend to asymptotically reach an equilibrium condition
with the binary being circularized \citep{1981A&A....99..126H, 2014MNRAS.444..542R}.

\begin{figure}
    \centering
    \includegraphics[width=\hsize]{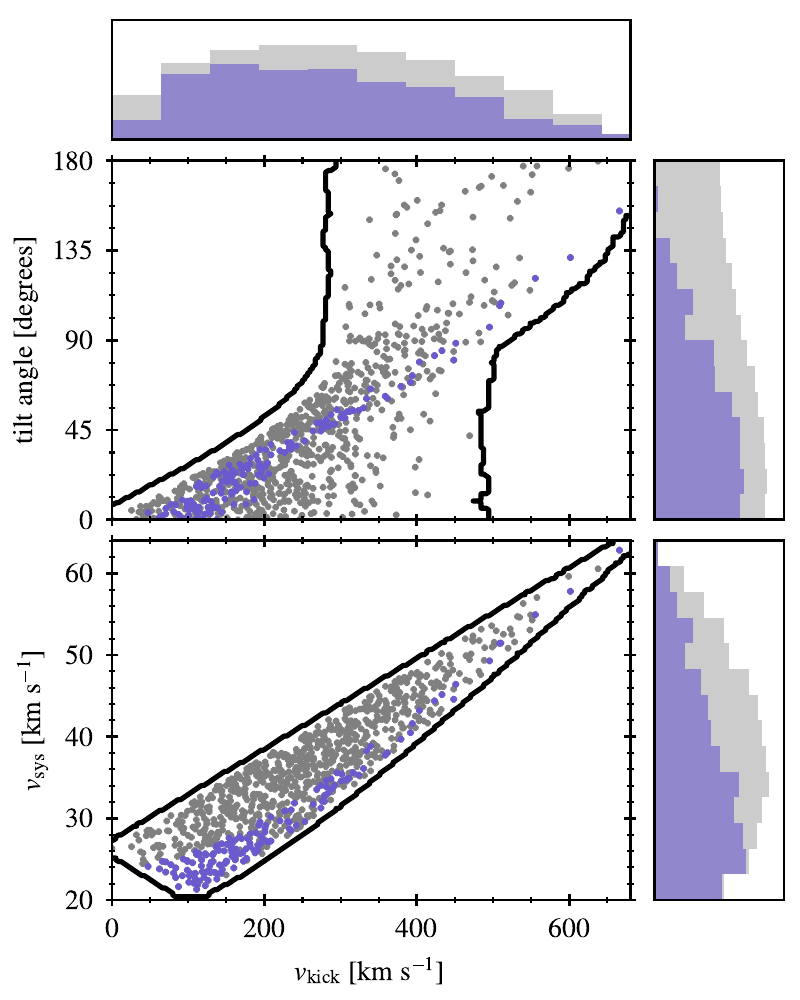}
    \caption{Post-kick inclination of the orbital plane (with respect to the pre-kick plane) and binary systemic velocity, $v_{\rm sys}$,
      as a function of the natal kick magnitude $v_{\rm kick}$. In light blue we show binaries matching the orbital parameters of
      GX~301--2 in its HMXB phase, while grey points represent all the natal kicks explored in detail. In the top and on the right, we show
      the weighted histograms. Black solid lines show the contour of the 99\% credible regions for the chosen kicks' distributions.}
    \label{fig:progenitors_kick_properties}
\end{figure}

Natal kicks for binaries that best match the orbital properties of GX~301--2 lie on a well-defined region for the kick velocity magnitude
and inclination between the pre- and post-kick orbit, as seen in Figure~\ref{fig:progenitors_kick_properties}. Increasing the magnitude of
the kick favours higher inclination values: for $v_{\rm kick} \gtrsim 450$~km~s$^{-1}$, we only obtain progenitors with a tilted post-SN
orbit $\gtrsim \pi/2$, suggesting that a retrograde NS might be reminiscent of the natal kick imposed during the explosion of the NS
progenitor \citep{1983ApJ...267..322H,1995MNRAS.274..461B,2000ApJ...541..319K}. An ongoing debate on the orientation of the spin of the NS
in GX~301--2 exists, with some authors favouring the scenario of a retrograde spinning NS based on data from the {\it Fermi}~Gamma-Ray
Burst Monitor \citep{2020MNRAS.494.2178M} and others favouring a prograde spin of the NS \citep{2020MNRAS.496.3991L}. In our simulations,
we have determined that only when the strength of the kick is above $450$~km~s$^{-1}$, does the binary contain a retrograde spinning NS.
One important caveat concerning this analysis comes from the fact that we do not consider stellar rotation. This implies that the role that
tides have in the evolution is not exhaustively taken into account. Tidal evolution tends to lead a binary to an equilibrium state of
coplanarity, circularity, and corotation \citep{1981A&A....99..126H}, thus by assuming non-rotating stars we are forcing coplanarity to be
reached instantly. Additionally, evolution until co-rotation can change the orbital angular momentum, impacting different orbital
parameters such as $P_{\rm orb}$.

We also find a clear trend for the systemic velocity ($v_{\rm sys}$) of the binary (see Figure~\ref{fig:progenitors_kick_properties}), in
which higher $v_{\rm kick}$ implies a higher post-kick $v_{\rm sys}$. A recent analysis of {\it Gaia} observations derived a value of
$v_{\rm sys} \sim 56$~km~s$^{-1}$ \citep{2022arXiv220603904F}, which in our simulations is matched for $v_{\rm kick} \gtrsim
450$~km~s$^{-1}$.

\begin{figure}
    \centering
    \includegraphics[width=\hsize]{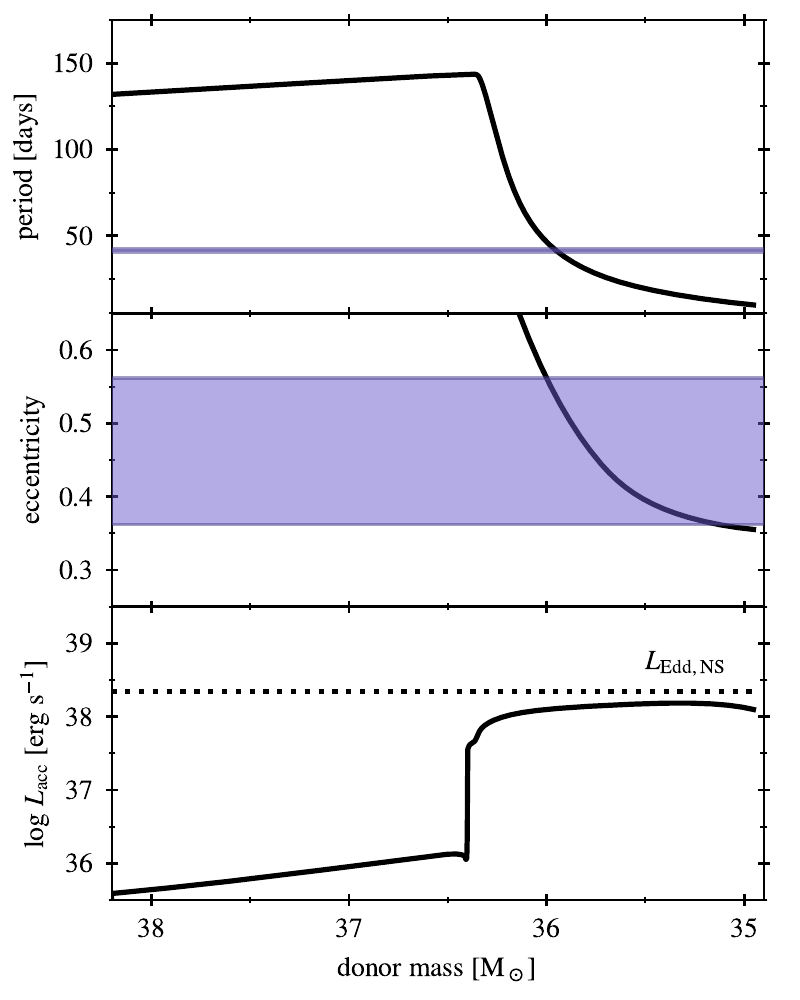}
    \caption{Evolution of orbital period, eccentricity, and released accretion energy of a selected binary model as a function of the mass
      of the optical companion. Time evolution proceeds from left to right. Light-blue regions represent observational constraints with
      increased error bars (see Section~\ref{section:kicks}). The x-axis corresponds to possible matching values for Wray~977, which is
      thought to be above~$33$~M$_\odot$ \citep{2006A&A...457..595K}.}
    \label{fig:progenitor_binary_properties}
\end{figure}

To illustrate the typical characteristics of the progenitors of GX~301--2, in Figure~\ref{fig:progenitor_binary_properties} we show the
evolution of the observational properties of one of the binaries, which has been selected as an example. We note that the range of donor
masses matching current measures of Wray~977 is thought to be larger than $33$~M$_\odot$. In the top panel of the figure, we show the
evolution of the orbital period after the formation of the NS. The orbital period initially increases due to strong winds from the
companion star. After leaving the main sequence, the star swells up and begins a fast MT phase in which the period decreases. Given the
high mass ratio, it is likely to result in an unstable MT phase followed by the development of a CE in which the NS would subsequently
merge with the core of the companion. During the phase in which the orbital period increases, the eccentricity remains almost constant as
there is no strong tidal circularization, as shown in the middle panel of Figure~\ref{fig:progenitor_binary_properties}. When the star
overflows its Roche lobe, the eccentricity begins to reduce and the system tends to circularize \citep{1995A&A...296..709V}.

We computed the luminosity released by accretion of matter~($L_{\rm acc}$) as follows:

\begin{equation}
    L_{\rm acc} = \dfrac{G M_{\rm NS} \dot{M}_{\rm acc}}{2 R_{\rm NS}}
,\end{equation}

\noindent where $G$ is the gravitational constant, $\dot{M}_{\rm acc}$ is the accretion rate onto the NS that considers the contribution
of wind accretion by the Bondi-Hoyle mechanism \citep{1944MNRAS.104..273B,2002MNRAS.329..897H}, and $R_{\rm NS}$ is the radius of the NS
that we set to $10$~km. As shown in Figure~\ref{fig:progenitor_binary_properties}, during the time leading up to the start of MT,
$L_{\rm acc}$ which was released by wind accretion is around $10^{36}$~erg~s$^{-1}$. Considering that a fraction of this energy is radiated
as X-rays, it can be detected by any current X-ray space telescope. After the MT starts, this luminosity reaches a maximum value of the
Eddington luminosity ($L_{\rm Edd,NS}$) as a direct consequence of having limited accretion up to the Eddington limit of the NS.


\begin{figure}
    \centering
    \includegraphics[width=\hsize]{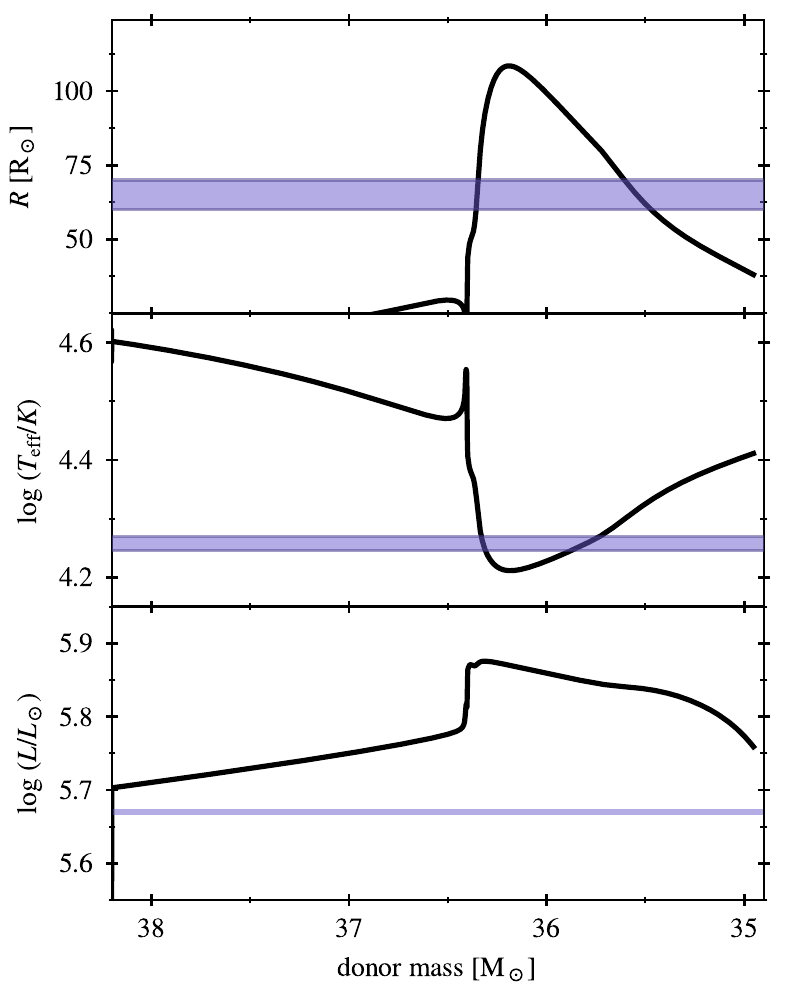}
    \caption{Evolution of the parameters of the donor star of the selected model. Time evolution proceeds from left to right. Light-blue
      regions represent observational constraints from \citet{2006A&A...457..595K}. In the case of the radius $R$, we use the range given
      by \citet{1999A&A...350..148W} of $60 < R < 70$~R$_\odot$. The x-axis range matches possible values for Wray~977, with a mass
      estimation of above $33$~M$_\odot$ \citep{2006A&A...457..595K}.}
    \label{fig:progenitor_donor_properties}
\end{figure}

Other important constraints are found in the behaviour of the companion star. As seen in Figure~\ref{fig:progenitor_donor_properties}, the
radius of the star begins too low to match the radius estimation from \citet{2006A&A...457..595K} of $R = 70$~R$_\odot$; also, just after
the star leaves the MS and travels through the HG, it is able to match the values quoted in \citet{2006A&A...457..595K} of
$R = 70$~R$_\odot$. This is also true for the effective temperature, $T_{\rm eff}$, of the star: it needs to be cooler than those (within
that mass range) found in the main sequence. On the other hand, the expected luminosity for the star is higher compared to the values
derived from observations of $\log\,L/L_\odot = 5.67$~\citep{2006A&A...457..595K}, which can be further studied by exploring the influence
that different free parameters of physical conditions in the stellar interiors -- such as convective overshooting, semi-convection
efficiency, among others -- have on the observable properties of a star
\citep{2012sse..book.....K,2020MNRAS.496.1967K,2020A&A...638A..55K}.


In order to compute the expected number of binaries with observational parameters measured for GX~301--2, we followed the method described
in \citet{2012MNRAS.419.2095M}, which provides an estimation for the number of HMXBs at any time as a function of the birth rate of compact
objects and the star formation rate (SFR). First, we derived the number of binaries with properties similar to those of GX~201--2 using the
birth rate of the NS in the system from the range of initial masses we obtained, $23 - 29$~M$_\odot$, assuming an initial mass distribution
as in \citet{1955ApJ...121..161S}. Also considering the constraints in the initial $P_{\rm orb}$ and $q$ and the natal kicks, we could then
derive the fraction that become X-ray sources such as GX~301--2 and its duration. We obtain a rate of $2.89 \times 10^{-3}$~per~SFR.
Additionally, assuming that the number of HMXBs with a luminosity $L > 10^{35}$~erg~s$^{-1}$ is $\approx 135$~per~SFR
\citep{2012MNRAS.419.2095M}, we determined that the fraction of HMXBs with properties similar to GX~301--2 is $6.64 \times 10^{-5}$,
implying a low chance of finding a similar binary in the Milky Way. For consistency, we used a different approach to compute the fraction
of HMXBs similar to GX~301--2, as presented in \citet{2018A&A...613L..10O}, and again obtain a very low fraction of $10^{-4}$.

\section{Discussion}
\label{section:discussion}

In this work, we have studied the progenitor properties and formation channel of the NS XRB GX~301--2. This HMXB is of particular
interest as it contains a NS accompanied by a very massive hypergiant star. In the standard picture of binary evolution, this system
requires the progenitor of the NS to have transferred most of its envelope to the companion, now known as Wray~977, in a highly
accretion-efficient scenario. We used the publicly available stellar-evolution code {\tt MESA} to evolve a grid of initial massive
binaries, starting from their position in the ZAMS until they reached the HMXB stage, including a natal kick distribution applied to the NS
that formed after one of the stars in the binary core collapsed. With the exploration of different strengths in the natal kicks, we were
able to find binary systems that match the particular and unique observational constraints associated with GX~301--2.

We found that only a very narrow range of initial masses have final masses compatible with GX~301--2. This narrow constraint in the initial
conditions is also true for the orbital period, as binaries with periods above $\sim$5~days contain stars with masses outside the derived
range, thus being unable to be compatible with GX~301--2. Given the highly uncertain nature of asymmetric natal kicks imparted when the NS
is born, kicks are generally treated in a stochastic way by assuming certain distributions for their strengths and directions, which
consequently lead to changes in important binary parameters such as eccentricity, separation, inclination between the spin of the
components and the orbital angular momentum, among others. Values for the strength of the natal kicks can be further constrained if new
observational data are considered: a value of $v_{\rm sys} \sim 56$~km~s$^{-1}$ \citep{2022arXiv220603904F} is only obtained for
$v_{\rm kick} \gtrsim 450$~km~s$^{-1}$. A kick velocity of such a magnitude produces a post-SN orbit with a tilt angle $\gtrsim \pi/2$,
which could be related to a retrograde NS present in this particular binary, as suggested by \citet[]{2020MNRAS.494.2178M}, although this
proposal is still under debate \citep{2020MNRAS.496.3991L}.

We caution that our range of preferred initial conditions is subject to several uncertainties due to the poorly constrained parameters
associated with the interior evolution of stars and to binary interactions. For instance, considering that rotating stars can have a large
impact on the range of values found, this means that it is likely that MT proceeds in a fully conservative way only until the accretor
reaches its critical rotational velocity, at which point accretion might be prevented by, for instance, the development of a strong wind
\citep{2005A&A...435.1013P, 2007A&A...465L..29C}. Additionally, the effects of tides after the formation of the NS are expected to have an
impact on how the angular momentum is redistributed within the binary, thus varying the spin of the companion star, the eccentricity, and
the orbital period \citep{1981A&A....99..126H, 2014MNRAS.444..542R}. In this work we use a Maxwellian distribution with a velocity
dispersion of $265$~km~s$^{-1}$ that was inferred from the proper motion of pulsars \citep{2005MNRAS.360..974H}. However, other works found
more complex distributions such as a bimodal distribution with characteristic velocities of $90$~km~s$^{-1}$ and $500$~km~s$^{-1}$ from the
velocities of isolated radio pulsars \citep{2002ApJ...568..289A}. There is also evidence of a low-kick population ($30$~km~s$^{-1}$)
and a high-kick population ($400$~km~s$^{-1}$) based on observed binary NSs \citep{2016MNRAS.456.4089B}. However, it is expected that
changes in the distribution of the strength of natal kicks used would not alter the distribution of orbital parameters compatible with
GX~301--2, but rather the probability of obtaining such configurations. 

We also study the expected fraction of binaries with properties matching those measured for GX~301--2, using, as priors, standard
distributions associated with the initial binary parameters and we found that the chances of producing such a binary are very low. However,
these results are tightly connected to the chosen SFR and the distribution of initial parameters; nevertheless, given the small region of
binary parameters leading to such a configuration, it is unlikely that changing these distributions would significantly increment the
number of expected GX~301--2-like binaries. 

\section{Conclusions}
\label{section:conclusions}

The unusual combination of masses measured in GX~301--2 makes it an interesting binary in which to study stellar evolution scenarios
involving highly efficient mass-accretion regimes between stars. To this end, we performed detailed binary evolution calculations in a wide
grid of initial parameters, which allowed us to find models with properties matching the ones measured in GX~301--2. Below, we summarize
the main results of this work:

\begin{itemize}
    \item {\bf Initial masses}: Possible progenitors, defined as systems that are able to form a binary with a NS orbiting around a star
      with a mass consistent with Wray~977, are only found for $M_1 = 23 - 30$~M$_\odot$ with companions such that the initial mass ratio
      is $q = 0.8 - 0.9$. Masses outside of this range either produce a BH as a compact object after the first collapse, or they contain a
      companion with a mass outside of the range derived for Wray~977.
    
    \item {\bf Orbital periods}: $P_{\rm orb} \lesssim 5$~days are favoured to produce such possible progenitors, thus leading to a Case A
      scenario of MT before the first SN event. Above this limit, the accretion is not sufficient enough to increase the mass of the
      companion to be inside the range derived for Wray~977. Additionally, $P_{\rm orb} < 2$~days are also disfavoured as the MT would end
      up producing a contact system.
    
    \item {\bf Asymmetric natal kicks}: Thanks to new astrometric observations, the natal kick strength is likely to have been
      $v_{\rm kick} \gtrsim 450$~km~s$^{-1}$, which is expected to produce a tilt angle of $\gtrsim \pi/2$.
    
    \item {\bf Estimated rates}: Weighting of our results with commonly used distributions of initial binary parameters, we estimate that
      the chances of having HMXBs with properties similar to the ones found in GX~301--2 are very low, that is less than two out of
      100\,000 binaries. We thus do not expect to find another binary with similar properties in the Milky Way.
\end{itemize}

Given the constraints derived from new observations available on the kinematics of different eccentric HMXBs, thanks to the extremely
precise astrometric measurements obtained with the {\it Gaia} satellite \citep{2021A&A...649A...1G} which allows for the systemic
velocity of these systems  to be inferred, incorporating more details as to binary evolutionary models can provide better constraints on
the evolution of progenitors until their current observed states as HMXBs. For GX~301--2, new observations of the source in different
electromagnetic bands will help to constrain the evolution even further, followed by its progenitor binary, its future prospects, and the
role it has on the environment. For instance, detailed measurements of surface abundances on Wray~977 would help to unravel the role of a
previous MT episode. It would be an interesting topic for a future work to incorporate the effects of rotation and tides to predict how
initial binary parameters are mapped to the parameters just after the formation of the NS until nowadays how GX~301--2 is observed.


\begin{acknowledgements}

ASB is a fellow of CONICET. FG and JAC are CONICET researchers. JAC is a Mar\'ia Zambrano researcher fellow funded by the European Union
-NextGenerationEU- (UJAR02MZ). FG and JAC acknowledge support by PIP 0113 (CONICET). This work received financial support from
PICT-2017-2865 (ANPCyT). JAC was also supported by grant PID2019-105510GB-C32/AEI/10.13039/501100011033 from the Agencia Estatal de
Investigaci\'on of the Spanish Ministerio de Ciencia, Innovaci\'on y Universidades, and by Consejer\'{\i}a de Econom\'{\i}a, Innovaci\'on,
Ciencia y Empleo of Junta de Andaluc\'{\i}a as research group FQM- 322, as well as FEDER funds. SC acknowledges the CNES (Centre National
d’Etudes Spatiales) for the funding of MINE (Multi-wavelength INTEGRAL Network). ASB, FG and SC are grateful to the LabEx UnivEarthS for
the funding of Interface project I10 « From binary evolution towards merging of compact objects ». Software:
{\tt MESA}\footnote{\url{http://mesa.sourceforge.net/}} \citep{Paxton2011, Paxton2013, Paxton2015, Paxton2018, Paxton2019},
{\tt ipython/jupyter} \citep{jupyter}, {\tt matplotlib} \citep{Hunter:2007}, {\tt numpy} \citep{harris2020array}, {\tt scipy}
\citep{2020SciPy-NMeth}.

\end{acknowledgements}

\bibliographystyle{aa}
\bibliography{aanda}

\end{document}